\documentclass[fleqn,usenatbib]{mnras}

\usepackage{newtxtext,newtxmath}

\usepackage[T1]{fontenc}
\usepackage{ae,aecompl}

\makeatletter
\newcommand{\@todonotes@enable}{1}
\newcommand{\@todonotes@inline}{1}
\makeatother
\input{notes.tex}

\usepackage{graphicx}	
\usepackage{amsmath}	
\usepackage{amssymb}	
\usepackage[normalem]{ulem}

\newcommand{\gaia}{$Gaia\,$}

\title[Solar Siblings]{Searching for Solar Siblings in APOGEE and \gaia\ DR2 with N-body Simulations}
\author[Webb et al.]{Jeremy J. Webb$^1$ \thanks{E-mail: webb@astro.utoronto.ca (JW), price-jones@astro.utoronto.ca (NPJ)}, Natalie Price-Jones$^1$, Jo Bovy$^1$, Simon Portegies Zwart$^2$, \newauthor Jason A. S. Hunt$^{3,4}$, J. Ted Mackereth$^{5,6}$ \& Henry W. Leung$^1$ 
 \\
$^1$Department of Astronomy and Astrophysics, University of Toronto, 50 St. George Street, Toronto, ON, M5S 3H4, Canada \\
$^2$ Leiden Observatory, Leiden University, P.O. Box 9513, 2300 RA, Leiden, The Netherlands \\
$^3$ Dunlap Institute for Astronomy and Astrophysics, University of Toronto, 50 St. George Street, Toronto, Ontario, M5S 3H4, Canada \\
$^4$ Center for Computational Astrophysics, Flatiron Institute, 162 5th Av., New York City, NY 10010, USA \\
$^5$Astrophysics Research Institute, Liverpool John Moores University, 146 Brownlow Hill, Liverpool, L3 5RF, UK \\
$^6$School of Astronomy and Astrophysics, University of Birmingham, Edgbaston, Birmimgham, B15 2TT, UK \\
}

\date{Accepted XXX. Received YYY; in original form ZZZ}

\pubyear{2019}

\begin{document}
\label{firstpage}
\pagerange{\pageref{firstpage}--\pageref{lastpage}}
\maketitle

\begin{abstract}


We make use of APOGEE and \gaia data to identify stars that are consistent with being born in the same association or star cluster as the Sun. We limit our analysis to stars that match solar abundances within their uncertainties, as they could have formed from the same Giant Molecular Cloud (GMC) as the Sun. We constrain the range of orbital actions that solar siblings can have with a suite of simulations of solar birth clusters evolved in static and time-dependent tidal fields. The static components of each galaxy model are the bulge, disk, and halo, while the various time-dependent components include a bar, spiral arms, and GMCs. In galaxy models without GMCs, simulated solar siblings all have $J_R < 122$ km $\rm s^{-1}$ kpc, $990 < L_z < 1986$ km $\rm s^{-1}$ kpc, and $0.15 < J_z < 0.58$ km $\rm s^{-1}$ kpc. Given the actions of stars in APOGEE and \gaia, we find 104 stars that fall within this range. One candidate in particular, Solar Sibling 1, has both chemistry and actions similar enough to the solar values that strong interactions with the bar or spiral arms are not required for it to be dynamically associated with the Sun. Adding GMCs to the potential can eject solar siblings out of the plane of the disk and increase their $J_z$, resulting in a final candidate list of 296 stars. The entire suite of simulations indicate that solar siblings should have $J_R < 122$ km $\rm s^{-1}$ kpc, $353 < L_z < 2110$ km $\rm s^{-1}$ kpc, and $J_z < 0.8$ km $\rm s^{-1}$ kpc. Given these criteria, it is most likely that the association or cluster that the Sun was born in has reached dissolution and is not the commonly cited open cluster M67.

\end{abstract}


\section{Introduction} \label{intro}

The Sun is easily the most well-studied star in the Universe, with its present day properties being very well constrained. The circumstances of the Sun's formation and that of our solar system, however, are still relatively unknown \citep{BlandHawthorn10,Adams10}. A firm understanding of how the Sun formed will pave the way for a more detailed knowledge of star formation in general, which in turn will help trace the star formation history of the Milky Way. Furthermore, a greater understanding of the formation of the only known solar system to harbour life is essential in the pursuit of searching for habitable worlds and life beyond our own planet. 

The properties of the giant molecular cloud from which the Sun formed have been particularly difficult to constrain as it is not easy to identify and confirm whether or not other stars in the Milky Way were born in the same environment (i.e. solar siblings) \citep[e.g][]{Brown10, Bobylev11, Batista12, Batista14, Liu15, Martinez16, Abolfathi2018}. It is also not entirely clear if the Sun's progenitor system was an unbound association of stars, a star cluster that has since fully dissolved, or exists in the form of an open cluster that has since migrated away from the Sun's current orbit \citep{Pichardo12, Gustafsson16, Jorgensen19}.

The ability of planetary systems to form and survive in star clusters is also of interest and has been used to constrain the properties of the Sun's birth cluster, as close encounters with solar siblings can lead to planets becoming unbound from their host star \citep[e.g.][]{PortegiesZwart15, vanElteren2019}. Taking into consideration the ability of surviving planetary systems to also form life, several studies suggest that the Sun's birth system was most likely a loosely bound or unbound association \citep{deJuanOvelar12,Winter19}.


In the cases of the Sun forming in an unbound association and the Sun's birth cluster having reached dissolution, \citet{PortegiesZwart09} finds that several solar siblings are still likely to be found within 100 pc of the Sun. In the case of the Sun escaping from a cluster that still exists today, M67 is often cited as the potential birthplace of the Sun because its members have solar-like abundances \citep{Heiter14,Onehag14}. However, whether or not M67 is spatially and kinematically related to the Sun is up for debate \citep{Pichardo12,Gustafsson16}. Most recently, \citet{Jorgensen19} showed that it is possible for the Sun to have escaped M67 if the cluster formed close to the Sun's current orbit and then quickly migrated away from the plane of the disc due to interactions with GMCs. This migration also increases the cluster's ability to survive as the frequency of damaging GMC interactions decreases \citep{Buckner2014} with height above the disc. Confirming solar siblings kinematically and measuring their abundances would constrain the properties of the solar birth environment \citep{BlandHawthorn10, Adams10}, and would limit the amount of enrichment that can occur within a given GMC during star formation. 

The sizes and densities of collapsing giant molecular clouds yield between several tens to several millions of stars \citep{Lada2003}. Turbulent mixing within the gas cloud and the short timescale of formation relative to stellar lifetimes means that stars in the same birth cluster will share the same chemical abundances and ages \citep{Feng2014}, although some intra-cluster enrichment may be possible if very massive stars form early \citep{BlandHawthorn10, Getman2018}. Only sufficiently dense populations, such as globular clusters, are able to survive as bound star clusters for more than a few gigayears and show evidence of multiple formation events \citep{Carretta2009}. Typically younger and smaller, open clusters are often taken as proxies for still intact birth clusters. Most low density birth clusters, despite forming as bound systems, will dissolve in less than 100 Myr \citep{Lada2003}. Unbound stellar associations, on the other-hand, break up almost immediately. Given that the Sun has clearly escaped its progenitor association or cluster, we can only identify solar siblings that have been dispersed in phase space by using chemical tagging, grouping stars based on their chemical abundances \citep{Freeman2002}. 

Chemical tagging relies on two major assumptions: first, that birth clusters are chemically homogeneous, and second, that each birth cluster has a sufficiently unique chemical signature as to make its members distinct from members of other birth clusters in chemical space. Despite the great promise of this technique, theoretical predictions have shown that it will be quite challenging \citep{Ting2015}. The assumptions needed for chemical tagging have been tested extensively with open clusters. Several studies have investigated the homogeneity of open clusters and found them to be quite chemically homogeneous (e.g. \citealt{DeSilva2006}, \citealt{DeSilva2007}, \citealt{Bovy2016}, \citealt{Price-Jones2018}). However, close studies of individual open clusters have also been able to identify discrepancies between the chemical signatures of member stars \citep{Liu2016a}, as well as identifying systematic differences between subgiant and main sequence turn off stars \citep{Liu2016}. These cases indicate the importance of considering lifetime evolution of surface abundances and required level of chemical homogeneity within a cluster when undertaking chemical tagging. However, if stars of single evolutionary stage are selected and the spread of abundances within a cluster remains at the level of measurement uncertainties in large surveys, open clusters do represent a single chemical population \citep{Price-Jones2018}.

Attempts at chemically tagging smaller populations (<1000 stars) have also highlighted challenges with assuming unique chemical signatures for each birth cluster (e.g. \citealt{Mitschang2014}, \citealt{Blanco-Cuaresma2015}). Analyses of large spectroscopic samples like \citet{Ness2018} have identified the possibility of false siblings. Blind attempts at chemical tagging have been successful in recovering globular clusters, with their more unique signatures \citep{Chen2018}, as well as previously unidentified groups \citep{Hogg2016}. More recent work has also successfully distinguished known open clusters \citep{Blanco-Cuaresma2018}, and this is line with expectations from simulations \citep{Price-Jones2019}. 


Rather than a blind search of chemical space for structure that might reflect birth clusters, we narrow our focus to stars with solar-like abundances, employing chemically tagging on a small scale.

The Sloan Digital Sky Survey's (SDSS - \citealt{Eisenstein2011}, \citealt{Blanton2017}) Apache Point Observatory Galactic Evolution Experiment (APOGEE - \citealt{Majewski2017}) is ideal for searching for potential siblings, as it has accurately measured a range of chemical abundances of hundreds of thousands of stars in the Milky Way. More recently, deep learning methods have been used to revise abundance measurements \citep{Leung2019} and distance estimates with the help of spectro-photometric distances from \gaia  DR2 \citep{Leung2019b}. Classifying stars within APOGEE as solar siblings based on abundances alone is, unfortunately, quite difficult. Within APOGEE DR14, there are over 19,000 stars with high signal noise (S/N $>$ 50) spectra that have solar values of [Fe/H] within their measurement uncertainties. Incorporating other elements into the solar sibling criteria allows this list to be trimmed to between approximately 3400 and 9500 stars, depending on which elements are used. However it cannot be determined whether these stars formed in the same cluster as the Sun or simply in a GMC with comparable chemical abundances. The fact that these stars are spread out across the plane of the sky and have a wide range of Galactocentric distances suggests the latter.

Several recent studies have shown that the kinematic properties of stars can also be used to help establish whether or not stars share a formation environment. Most recently, simulations by \citet{Kamdar19} demonstrated that co-moving disc stars that are spatially ($\delta r < 20$ pc) and kinematically ($\delta v < 1.5$ km/s) near to each other likely formed from the same GMC. The authors followed up this work by identifying over a hundred nearby co-moving stellar pairs that could each have formed in the same progenitor cluster. With respect to solar siblings, earlier work by \citet{Martinez16} generated a suite of simulations of star clusters on solar orbits in a range of galaxy models and found that most simulated solar siblings fall in a narrow range of sky position and proper motion. Applying this criteria to a list of previously established solar sibling candidates \citep[e.g][]{Brown10, Bobylev11, Batista12, Batista14, Liu15}, \citet{Martinez16} finds that of the candidates with positions and kinematics that are consistent with their simulated solar siblings, previous estimates of their chemical abundances suggest they have a low probability of being a solar sibling. Hence combining chemical tagging with stellar kinematics is required to confirm that two or more stars formed in the same GMC. The combined datasets of APOGEE and the most recent \gaia data release (DR2), however, has yet to be used in the search for solar siblings and offers the largest dataset of stars with measured abundances, 3D positions and 3D velocities.


The purpose of this study is to search for solar siblings in the subset of APOGEE stars that have had their proper motions measured by \gaia with the help of an analysis of how different components of the Milky Way affect the distribution of stars that escape star clusters. The chemical tagging of APOGEE stars allows for an initial candidate list to be generated, which we then compare to $N$-body simulations of star clusters that end up on solar orbits. Motivated by \citet{Martinez16}, we consider star clusters that quickly reach dissolution in a static galaxy model containing a bulge, disc, and halo. Since the cluster dissolves so quickly, escaping stars will have similar distributions to models where the Sun was born in an unbound association as most stars are unaffected by interactions within the cluster. We also consider the effects of a bar, spiral arms, and GMCs on the spatial and kinematic properties of stars in each simulation. Each of these factors have been shown to affect cluster evolution, specifically with respect to stellar escape \citep{Gnedin1999, Gieles2006, Gieles07, Fujii12, Martinez16, Rossi2018, Jorgensen19, Mackereth19}. The simulations allow us to then determine the range of actions solar siblings are likely to have, which can then be applied to stars in APOGEE with solar abundances.

In Section \ref{s_observations} of this paper we introduce and summarize the APOGEE dataset and in Section \ref{s_simulations} we outline the simulations used to constrain the actions of potential solar siblings. In Section \ref{s_results} we first present our initial solar sibling candidate list based on chemical tagging alone. We then discuss the simulations of star clusters dissolving in both static and time dependent galaxy models and explore the range of actions that solar sibling stars may have. In Section \ref{s_discussion} we apply the results of our simulations to the APOGEE dataset in order to generate a list of stars that have solar abundances and orbits which agree overall with our star cluster simulations. APOGEE stars with actions closer to the action distribution of stars that escape simulated star clusters in static tidal fields are considered to have high probabilities of being solar siblings. APOGEE stars that overlap with our simulations of star clusters but are farther from the static potential case are considered lower probability candidates as their association with the Sun is dependent on the specific properties of the Galactic bar, spiral arms, and the Galaxy's GMC population. The complete list of APOGEE stars with solar abundances and kinematic properties that are aligned with our entire suite of simulations are provided in an on-line catalogue\footnote{\textbf{https://doi.org/10.5281/zenodo.3723953}}. We also discuss the possibility of M67 being the Sun's birth cluster given the cluster's actions. In Section \ref{s_conclusion} we summarize our findings.

\section{Data} \label{s_observations}

In order to firmly constrain whether or not a given star is a potential sibling of the Sun or not, we require knowledge of the star's metallicity, key elemental abundances, and the star's six dimensional spatial and kinematic properties. The latter criteria is necessary in order to solve the star's orbit and calculate its actions, assuming a given Galactic potential. Cross-matching the APOGEE and \gaia DR2 \citep{Gaia2018} catalogues provides a data set of stars with the necessary information. Comparing the measured actions to those of the Sun and simulations of star clusters on Sun-like orbits will help further constrain whether stars with similar abundances to the Sun could potentially have formed in the same birth cluster. Ideally one would also require a star's age to be 4.65 Gyr within uncertainty in order to be considered a solar sibling. Unfortunately the mean uncertainty of APOGEE star ages is $\sim$ 2 Gyr \citep{Mackereth19}, which is too high to strongly argue that a given star was born at the same time as the Sun or not. 

Chemical abundances for the stars in our sample are taken from the APOGEE survey \citep{Majewski2017}. This high resolution (R$\sim$22,500) survey observes stars with a H-band (1.5$\mu$m - 1.7$\mu$m) spectrometer mounted on the Apache Point Observatory 2.5 m telescope \citep{Gunn2006}. The survey has primarily targeted giant stars in fields across the Milky Way's disc and halo \citep{Zasowski2013, Zasowski2017}, with a sample size of ~250,000 stars in data release 14 \citep{Abolfathi2018, Holtzman18}.

Stars observed by APOGEE-2 are assigned chemical abundances by the APOGEE Stellar Parameter and Chemical Abundances Pipeline (ASPCAP - \citealt{GarciaPerez2016}). However in this work we make use of abundances estimated from the spectra by the \texttt{astroNN} deep learning package (\citealt{Leung2019} - \url{ https://github.com/henrysky/astroNN}), which was trained on the results of ASPCAP but is significantly faster and obtains higher precision abundances than ASPCAP even when the signal to noise ratio of a spectrum is below APOGEE's target of 100.


To determine what constitutes solar abundances in APOGEE, we make use of stars that are known to be members of the open cluster M67. M67 has been shown to have solar abundances when observed in other wavelength bands, \citep{Heiter14, Onehag14}, so we treat the mean abundances of stars in the cluster as our solar baseline for APOGEE. The mean abundances of stars in M67 and the standard error in the mean are listed in Table \ref{table:solarabundances} for the elements considered in this study.

\begin{table}
\centering
\begin{tabular}{lccc}
\hline
Element & Abundance & Error \\
\hline
{[Fe/H]} & 0.066 & 0.009 \\
{[Mg/Fe]} & 0.010 & 0.002 \\
{[Al/Fe]} & -0.012 & 0.007 \\
{[Si/Fe]} & -0.011 & 0.004 \\
{[K/Fe]} & -0.013 & 0.004 \\
{[Ca/Fe]} & -0.008 & 0.003 \\
{[Ni/Fe]} & 0.015 & 0.002 \\

\hline 

\end{tabular}
\caption{\label{table:solarabundances} Mean abundance of stars in M67 for select elements.}
\end{table}


Spatial and kinematic information for each star, which are necessary to solve their orbits and calculate their actions, are taken from APOGEE and \gaia DR2 \citep{Gaia2018}. On-sky positions (RA, Dec) and line-of-sight velocities are taken directly from APOGEE and proper motions are taken from \gaia DR2. We make use of stellar distances as calculated by \citet{Leung2019b}, who use the \texttt{astroNN} deep learning package to estimate distances using both spectra from APOGEE and photometry from 2MASS \citep{Skrutskie2006}, by training on \gaia parallax data.

It is important to note that in our search for solar sibling candidates, we assume that the surface abundances measured today are representative of the initial abundances at formation. Since we specifically focus on giant stars in the APOGEE catalogue, it is likely the case that some abundances have changed since formation. For example, surface levels of C, N, and O are known to change as stars evolve along the red giant branch during `dredge-up'. This convective mixing modifies the C/N ratio, and that quantity can be used to predict stellar age \citep{Martig2016}. Therefore we do not require the C, N, and O values of stars in APOGEE to be near-solar.

Atomic diffusion is another internal process that can change a star's surface abundances over its lifetime. The term encompasses a variety of gradient-driven processes that can manipulate surface chemistry. \citet{Dotter2017} found that while atomic diffusion can change surface abundances over a star's lifetime, once it has reached the giant stage the surface abundances of elements affected by atomic diffusion reflect their initial abundances. The influence of atomic diffusion has been observed by \citet{Souto2019} and \citet{Liu2019} in M67, where stars in different evolutionary stages were found to have differing abundances. However, in agreement with \citet{Dotter2017}, both works found that the giant stars in M67 exhibited little spread in abundance. Given that stars in APOGEE are primarily giant stars, atomic diffusion will not be a factor when tagging stars as potential siblings based on their abundances.

In addition to these internal processes, external processes may also influence a star's surface abundances; mass transfer from a companion or accretion of an orbiting body might have short or long term impacts on the star's chemistry, depending on the amount of mass involved. These effects have been invoked as a possible explanation for differences in chemistry between solar twins (\citealt{Melendez2009}, \citealt{Ramirez2010}) and stars in the same open cluster (e.g. \citealt{Liu2016}). 
 
Our selection of solar siblings based on their measured abundances does not take into consideration these sources of possible surface abundance evolution. Hence, we may be including some stars that were not born with abundances similar to the Sun but whose surface chemistry has become more solar over time. However, we expect that most members of the same birth clusters should share abundances within their measurement uncertainties, as open clusters are found to be overall chemically homogeneous \citep{Bovy2016}. In addition, the inclusion of the actions as a way to constrain a star's similarity to the Sun adds further confirmation that the star was likely born in an environment similar to the Sun's birth environment (so as to be put onto a similar orbit).

\section{Simulations} \label{s_simulations}

To simulate the evolution of star clusters on solar-like orbits, we make use of the Barnes \& Hut Tree code (BHTREE; \citealt{Barnes86}) made available through the Astrophysical Multipurpose Software Environment (AMUSE; \citealt{PortegiesZwart13, Pelupessy2013,PortegiesZwart2018}). Clusters were initialized as Plummer models with masses and half-mass radii of either 510 $M_{\odot}$ and 0.5 pc or 804 $M_{\odot}$ and 3 pc. This set of initial conditions marks the most and least dense model clusters that \citet{Martinez16} find could represent the Sun's birth cluster. It is necessary to include both the low and high density models as the spatial distribution of escaped stars can be quite different \citep{Martinez16}. 

We assume four different analytic models for the background Galactic potential in order to account for differences in cluster evolution between static and time-dependent tidal fields. In all cases, the underlying static tidal field is the \texttt{MWPotential2014} model from \citet{Bovy15}. In this model, the Milky Way consists of a bulge that is represented by a spherical power-law potential, a \citet{Miyamoto75} disc, and a \citet{Navarro96} halo. The three time-dependent cases we consider add either a rotating central bar or a rotating central bar with two different types of spiral arms to the \texttt{MWPotential2014} potential. The rotating bar model, taken from \citet{Dehnen2000} (see also \citet{Hunt18}), is assumed to have already formed and grown to its current state at the start of the simulation. This bar has a pattern speed of 35.75 km $\rm s^{-1}$ $\rm kpc^{-1}$, consistent with recent observations \citep{Portail2017, Sanders2019, Bovy2019}. Individual spiral arms are taken to be the \citet{Cox02} sinusoidal potential, with galaxy models consisting of either 2 or 4 spiral arms modelled as either a density wave or transient arms. The density wave spiral arms have a rotation speed of 21.725 km $\rm s^{-1}$ $\rm kpc^{-1}$ while the transient spiral arms rotation is radially dependent and corresponds to the rotation curve of the galaxy (see \citet{Hunt19} for a complete description of these potentials).

The properties of each Galaxy component can be found in Table \ref{table:galaxy_models}. Note that for the Galactic Bar and Density Wave Spiral Arm, $\phi_i$ and $\phi_{ref}$ are the present day values of Galactocentric $\phi$. In the transient spiral case, over the duration of each simulation we allow new arms to grow as old arms decay that there are always two active spiral arms in the galaxy model at a given time. A transient arm is set to reach its maximum strength every 0.46 Gyr Myr starting from 1 Gyr before the simulation starts to present day, with transient arm lifetimes also equaling 0.46 Gyr. It is important to note that in Table \ref{table:galaxy_models}, $\phi_{ref}$, along with $M_{arm}$ and $\alpha_{arm}$, refer to the properties of the arms when their strength is at a maximum. 

The tidal fields were easily incorporated into AMUSE via a new potential function in the galactic dynamics software package \texttt{galpy} \footnote{http://github.com/jobovy/galpy} \citep{Bovy15} that allows for any potential in \texttt{galpy} to be used in AMUSE.

\begin{table}
\centering
\begin{tabular}{lccc}
\hline
Potential  & Variable & Value \\
\hline
{Static Potential} \\
\hline 
{Bulge} \\
{} & {$M_b$} & {2.7e9 $M_{\odot}$} \\
{} & {$\alpha_b$} & {1.8} \\
{Disc} \\
{} & {$M_d$} & {6.8e10 $M_{\odot}$} \\
{} & {$a_d$} & {3 kpc} \\
{} & {$b_d$} & {0.28 kpc} \\
{Halo} \\
{} & {$M_h$} & {4.4e11 $M_{\odot}$} \\
{} & {$a_h$} & {16 kpc} \\
\hline
{Galactic Bar} \\
{} & {$A_{\rm f,bar}$} & {645.3 km$^{2}$ s$^{-2}$} \\
{} & {$r_{\rm bar}$} & {5 kpc} \\
{} & {$\Omega_{\rm bar}$} & {35.75 km s$^{-1}$ kpc$^{-1}$} \\
{} & {$\phi_i$} & {$25^{\circ}$} \\
\hline
{Density Wave Spiral Arm} \\
\hline 
{} & {Number of Arms} & {4} \\
{} & {$\rho_{0,\rm arm}$} & {0.18 $M_{\odot} \ pc^{-3}$} \\
{} & {$\Omega_{\rm arm}$} & {21.725 km s$^{-1}$ kpc$^{-1}$} \\
{} & {$\alpha_{\rm arm}$} & {$12^{\circ}$} \\
{} & {$\phi_{\rm ref}$} & {$45^{\circ}$} \\ 

\hline
{Transient Spiral Arm} \\
\hline 
{} & {Number of Arms} & {2} \\
{} & {$\rho_{0,\rm arm}$} & {0.13 $M_{\odot} \ pc^{-3}$} \\
{} & {$\alpha_{\rm arm}$} & {$25^{\circ}$} \\
{} & {$\phi_{\rm ref}$} & {$25^{\circ}$} \\ 
{} & {Life Time} & {0.46 Gyr} \\

\end{tabular}
\caption{\label{table:galaxy_models} Properties of the static potential and components of the time dependent potentials in which star clusters are simulated.}
\end{table}

Assuming the Sun is currently located at a Galactocentric position of $x,y,z = 8.0, 0.0, 0.025 \mathrm{kpc}$ with velocity $vx,vy,vz = -11.1, 232.24, 7.25 \mathrm{km s^{-1}}$ \citep{Schonrich2010,Bovy2012}, the Sun's orbit is then integrated back 5 Gyr in each potential to determine its location in a given galaxy model at birth. Orbit integration is performed using \texttt{galpy}. We then initialize a model cluster at this location. The cluster is then evolved forward for 5 Gyr in 0.24 Myr increments using a softening length of 3 pc and a tree opening angle of 0.6 radians, after which the actions of all stars are calculated. It is important to note that regardless of the potential in which the cluster was simulated, actions are calculated assuming the external tidal field is simply \texttt{MWPotential2014}. Calculating actions in non-axisymmetric potentials are approximations at best and far from trivial.

Finally, we consider an additional set of galaxy models that are identical to the ones discussed above except that they contain GMCs. As previously discussed, GMCs have been shown to have a strong effect on cluster dissolution \citep{Gieles2006, Kruijssen11} and the orbital distribution of escaped stars relative to the progenitor cluster \citep{Martinez16,Jorgensen19}. Motivated by \citet{Banik19}, the initial distribution of GMCs is setup using a near-complete set of GMC masses, radii and positions from \citet{Miville17}. The catalogue consists of over 8000 GMCs that have a mean radius of 30 pc and surface densities between 2 and 300 $M_{\odot} \rm pc^{-2}$, with inner GMCs having higher surface densities than outer GMCs. Given that the catalogue is most complete locally, with clear incomplete patches in the outer disk and on the other side of the Galactic centre, we take the three dimensional cylindrical coordinates of all local GMCs within Galactocentric $\phi = \pm 45^{\circ}$ of the Sun and create four realizations of each GMC with a new random $\phi$. This approach effectively creates a uniform GMC population within our galaxy models. 

The orbits of the GMCs are set to be circular in \texttt{MWPotential2014} and then evolve in the same galaxy models as each host cluster. The forces acting on cluster stars due to the population of GMCs are calculated using the same BHTREE formalism as cluster star interactions (via the AMUSE bridge technique), with each GMC represented by an analytic Plummer sphere that has a softening length set equal to the GMC's radius. Given that the main goal of these additional simulations is to determine how GMCs can change the action distribution of solar siblings, we do not include a prescription for mass evolution of GMCs or the formation of new GMCs. Similarly, the orbits of GMCs are assumed to be unaffected by either the star cluster or other GMCs in the simulation. Hence the long-lived GMCs in our simulations act as a proxy for a random distribution of short-lived GMCS.

It should be noted that the centre of mass of each model stellar system will, by design, be located at the Sun's present day location after 5 Gyr of evolution. Hence the distribution of allowable solar sibling actions will always be centred around the Sun's current actions. Combined with our simplified model for including GMCs in the Galaxy potential, using the simulated distributions to find solar sibling candidates will be a conservative estimate, as it is more likely that the centre of mass of the Sun's birth cluster does not have a solar orbit and clusters undergo a wider range of GMC interactions. However a very large suite of simulations, one that explores a range of bar, arm and GMC properties, would be required to find all possible dissolving clusters that could produce a star with a Sun like orbit.  For the purposes of this study, we will focus on how different elements of the Milky Way affect the distribution of solar siblings and stress that our method for identifying solar siblings is conservative. 



\section{Results}\label{s_results}

With the combined APOGEE/\gaia datasets and our suite of simulations, we have all the necessary ingredients to search for solar sibling candidates. In the following we first generate a list of solar sibling candidates by chemically tagging stars in APOGEE with solar abundances. We then calculate the actions of each star in APOGEE with the help of proper motions from \gaia. The results of our $N$-body simulations are then presented to constrain the range of actions one can reasonably expect solar siblings to have.

\subsection{Abundances and Actions of Solar-like Stars in APOGEE and \gaia}

\begin{figure}
    \includegraphics[width=0.48\textwidth]{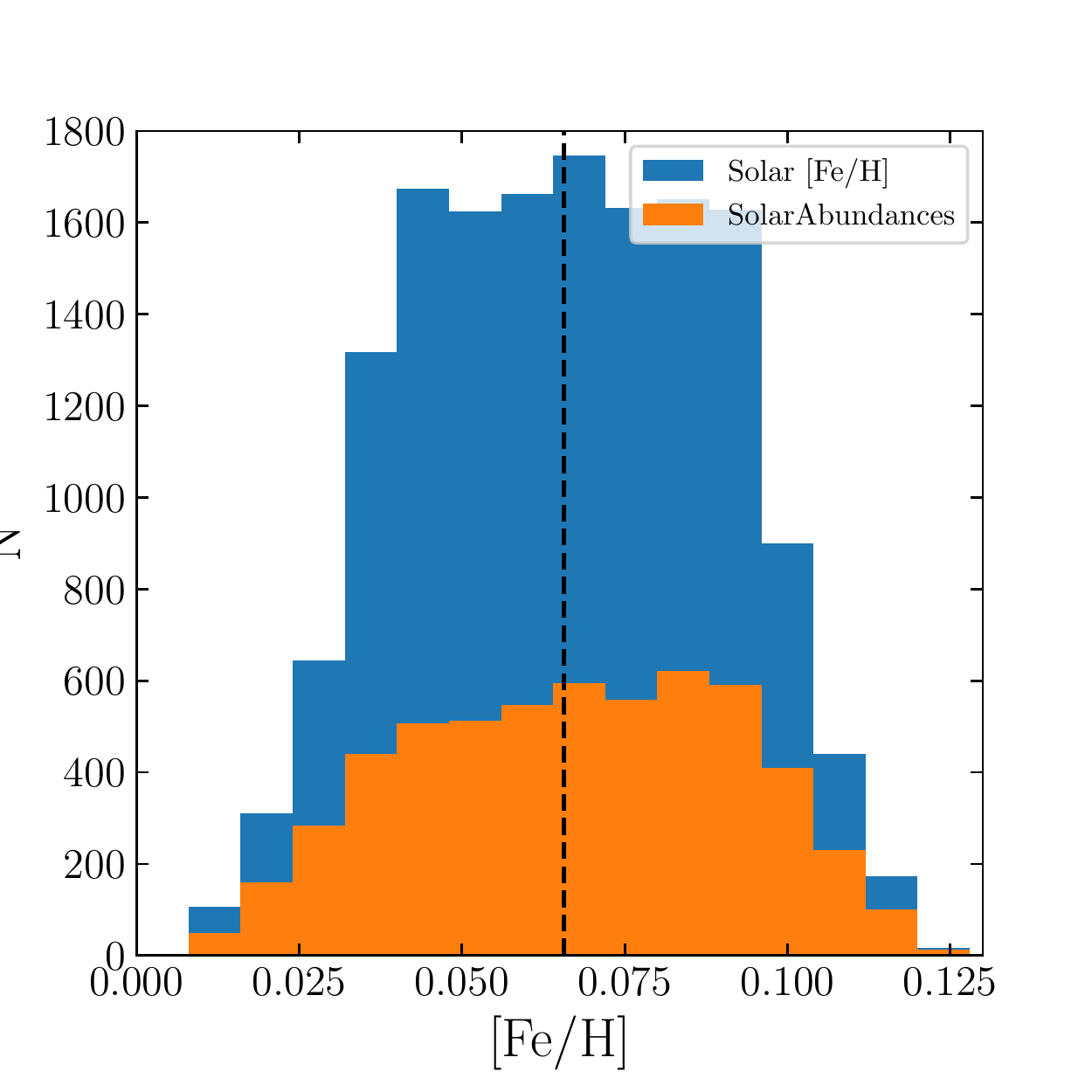}
    \caption{[Fe/H] distribution of APOGEE stars that are consistent with having metallicities that are solar within the measurement uncertainty (blue). The distribution of APOGEE stars that also have solar [Mg/Fe],[Al/Fe],[Si/Fe],[K/Fe],[Ca/Fe] and [Ni/Fe] abundances are shown in orange. The vertical dashed line marks the average [Fe/H] of stars in M67, which we take to be solar and is equal to 0.066.}
    \label{fig:apogee_feh}
\end{figure}

With updated APOGEE stellar abundances from the \texttt{astroNN} deep learning package, we limit our initial analysis to the 124,918 stars that have high signal to noise ratios (SNR > 50), uncertainties in log g less than 0.2, maximum uncertainties in [Fe/H] of 0.05, and maximum uncertainties in $T_eff$ of 20$\%$. These additional criteria result in our search for siblings being limited to stars with 0.5 $<\log g/<$3.7, and 3750$<T_{\rm eff}<$5330, while ensuring that we are only considering giant stars in APOGEE with high quality abundance measurements. Of this initial subset of stars, we find 10,429 stars have solar metalicities ([Fe/H]) within error. Figure \ref{fig:apogee_feh} illustrates in blue the distribution of stars with near-solar values of [Fe/H]. The dataset can be limited further to 3094 (orange histogram in Figure \ref{fig:apogee_feh}) by considering stars that also have solar-like values of [Mg/Fe], [Al/Fe], [Si/Fe], [K/Fe], [Ca/Fe], [Ni/Fe]. These elements have been shown to vary within star clusters by less than $\approx0.03\,\mathrm{dex}$ \citep{Bovy2016}, which is smaller than their mean cluster-to-cluster variation.


\begin{figure}
    \includegraphics[width=0.48\textwidth]{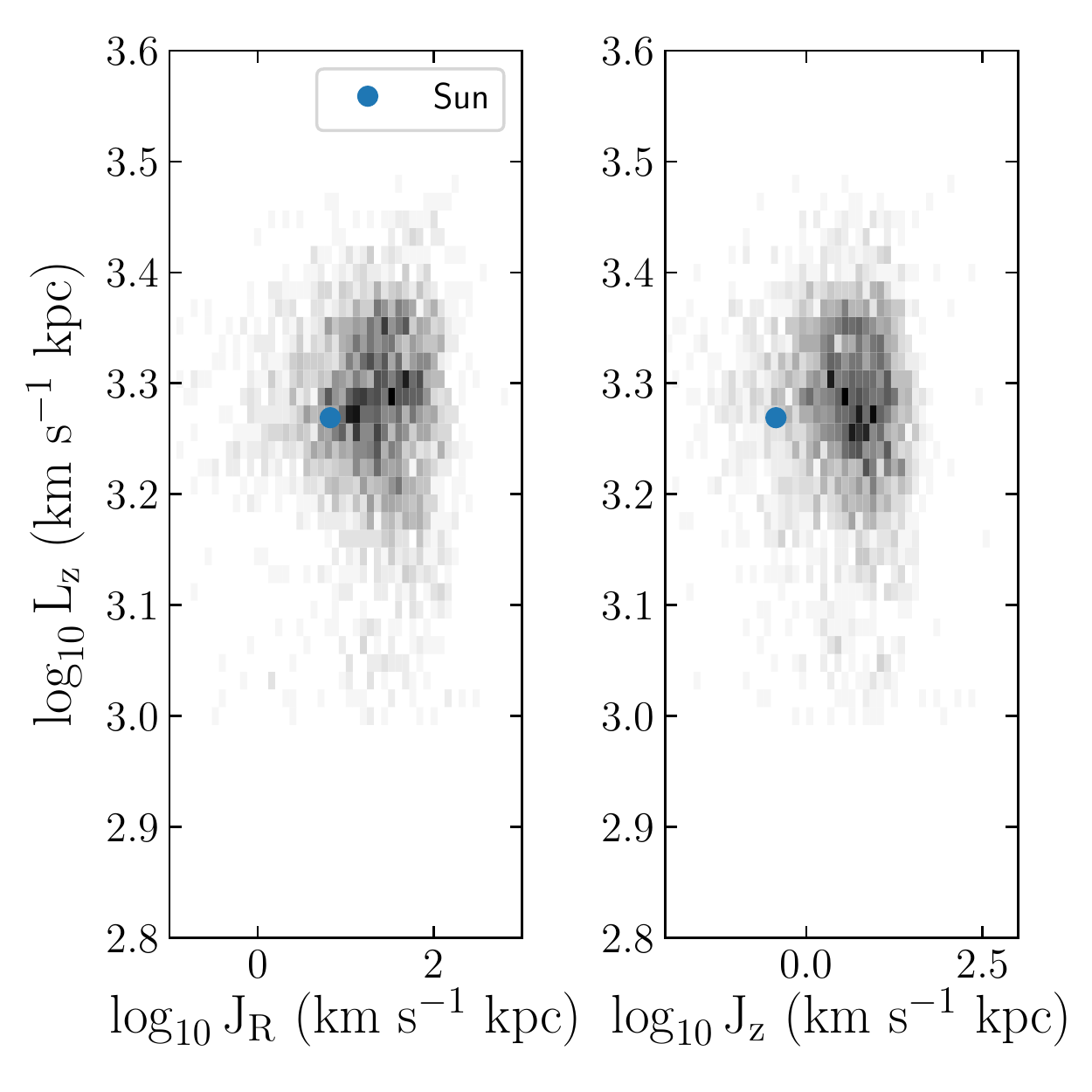}
    \caption{Distribution of actions $J_R$, $L_z$, and $J_z$ for APOGEE stars with solar abundances. The blue points mark the actions of the Sun.}
    \label{fig:apogee_actions}
\end{figure}

As previously discussed, it is impossible to conclude whether any of these $\sim$ 3000 stars actually formed in the same GMC as the Sun based on their abundances alone. We therefore calculate the actions $J_R$, $L_z$, and $J_{z}$ of each of the APOGEE stars with solar abundances as stars which escape a cluster will have comparable actions to their previous host. Actions are calculated using \texttt{galpy} assuming the \texttt{MWPotential2014} galaxy model. The $J_R$ - $L_z$, and $J_{z}$ - $L_{z}$ distributions of our subset of APOGEE stars are shown in Figure \ref{fig:apogee_actions}, with the Sun's actions marked in blue.

From Figure \ref{fig:apogee_actions} we see that the Sun lies outside the locus of points in the $J_R$ - $L_z$, and $J_{z}$ - $L_{z}$ parameter space. Hence most APOGEE stars with solar abundances lie on drastically different orbits than the Sun. Only a small subset of stars simultaneously have near-solar abundances \textit{and} actions. However as previous work by \citet{Martinez16} and \cite{Jorgensen19} have shown, it is possible for the orbit of the Sun's progenitor cluster to have changed between its time of formation and its present day location (if it hasn't reach dissolution). Hence the actions of solar-siblings will likely be spread out in a distribution around the actions of the Sun. To get a sense of what that distribution should be, we turn to $N$-body simulations.

\subsection{Positions and Actions of Solar Siblings in $N$-body Simulations}

To better constrain the probability that a star with solar-like abundances could have formed in the same progenitor cluster as the Sun, we turn to N-body simulations of stars clusters on solar orbits. From these simulations, it is possible to determine the range of actions that solar siblings will have. We begin our analysis by considering clusters evolving in galaxy models without GMCs, and then compare those results to simulations of clusters in identical galaxy models that do include GMCs.

\subsubsection{Solar Siblings in Analytic Potentials without Giant Molecular Clouds} \label{s_pots}

\begin{figure}
    \includegraphics[width=0.48\textwidth]{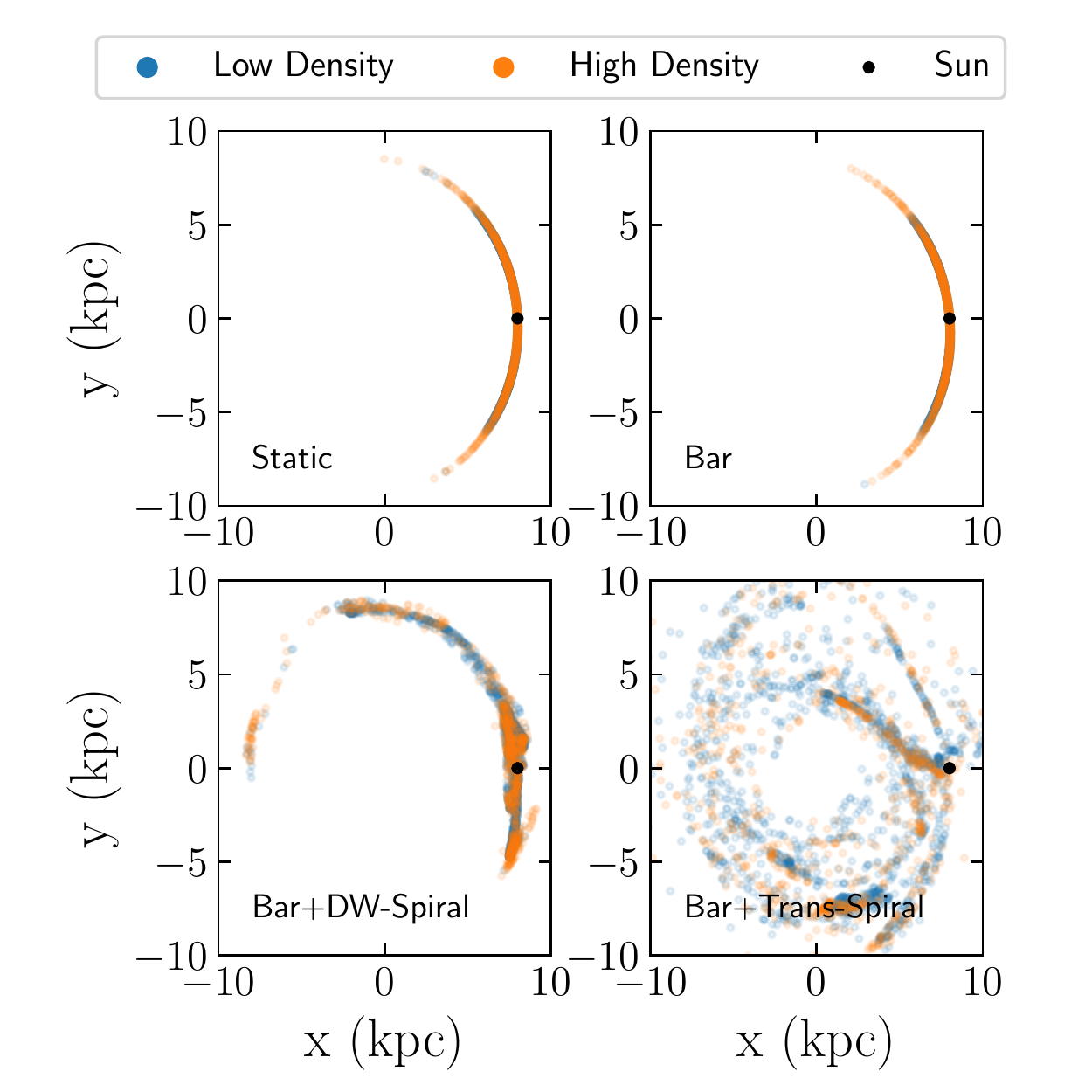}
    \caption{Spatial distribution of stars in low and high density model clusters orbiting in the static \texttt{MWPotential2014} tidal field (top left panel) and in time dependent tidal fields featuring a bar (top right panel), a bar with a density-wave spiral arm (bottom left panel) and a bar with transient spiral arms (bottom right panel). In each panel the black point marks the position of the Sun. While the bar has little effect on the dissolution of possible solar birth clusters, spiral waves lead to debris that is significantly more spread out.}
    \label{fig:amuse_positions}
\end{figure}

The present day spatial distributions of stars in each of our model clusters are shown in Figure \ref{fig:amuse_positions}. Each panel includes the distribution of stars that form inside clusters that have initially low (blue) and high (orange) densities. The four panels correspond to the static galaxy model (top-left), the static galaxy model with a bar (top-right), the static galaxy model with a bar and density wave spiral arms (bottom-left) and finally the static galaxy model with a bar and transient spiral arms. In each panel the Sun's present position is shown in black. The final distribution in $z$ is not shown as it remains fairly narrow for all four galaxy models. The orbit of each model cluster's centre of mass in the static and bar potentials, along with escaped stars themselves, stay within $\pm 0.1$ kpc of the Galactic plane. Stars orbiting in potentials containing the density-wave and transient spiral arms are pushed a little farther away from the Galactic plane, with stars distributed between $\pm 0.15$ kpc.

From Figure \ref{fig:amuse_positions} it can be seen that the presence of a bar alone does little to accelerate the dissolution of either model cluster. Spiral arms on the other hand have a very strong effect on cluster evolution. The density wave spiral arms cause periodic episodes of enhanced mass loss, which correspond to gaps in the stellar stream. They additionally appear to widen the distribution of velocities stars have as they escape the cluster and can further affect the orbits of stars after they have escaped the cluster, as the resultant stream is much thicker than the static potential case and contains spurs. 

Transient spiral arms are very destructive, as can be seen in the bottom-right panel of Figure \ref{fig:amuse_positions}. Only a few short streams are visible that would indicate these stars were once part of the same host cluster. Without kinematic \textit{and} chemical information, it would be impossible to conclude that stars outside of these short streams are solar siblings.

\begin{figure}
    \includegraphics[width=0.48\textwidth]{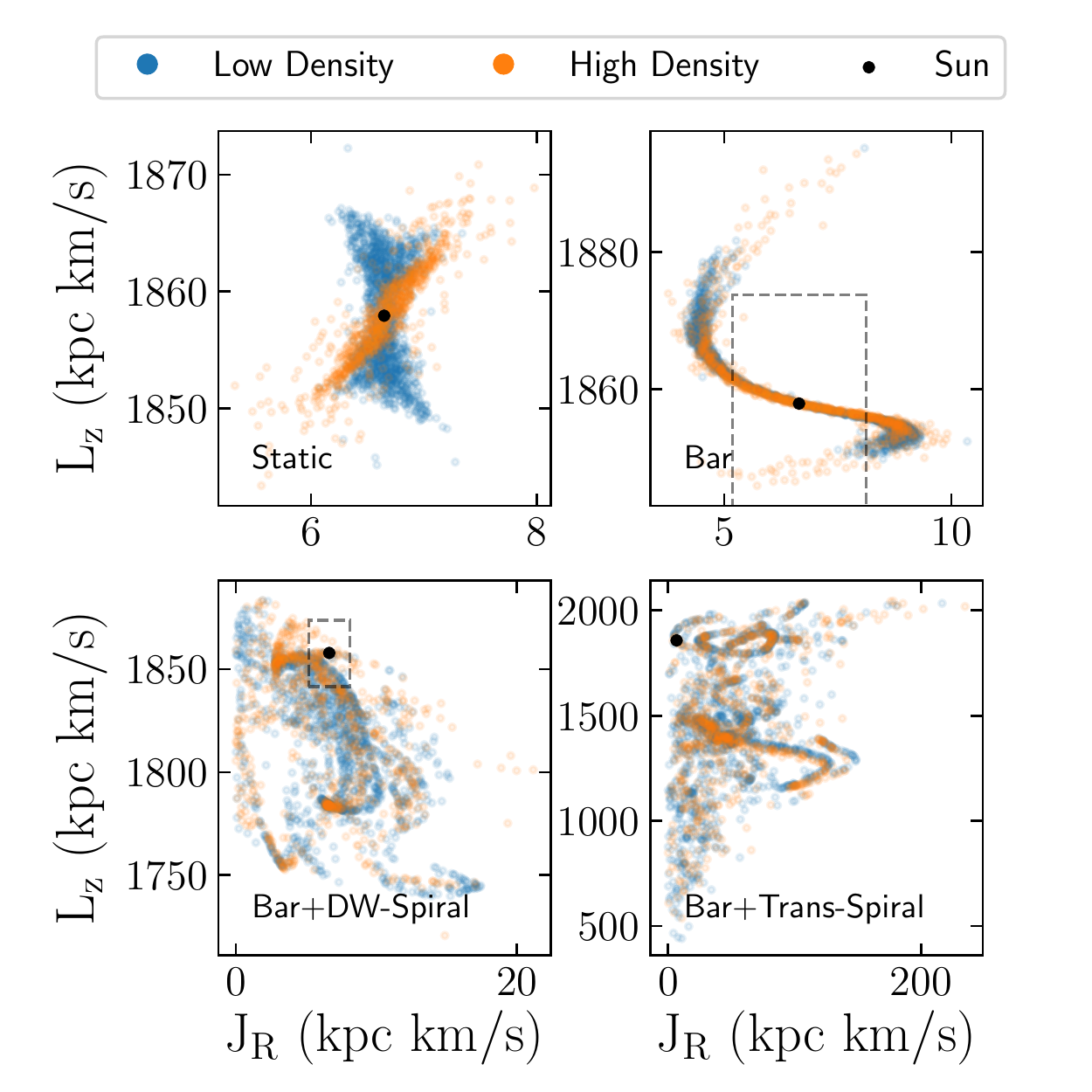}
    \caption{Action distribution ($J_R$ and $L_Z$) of stars in low and high density model clusters orbiting in the static \texttt{MWPotential2014} tidal field (top left panel) and in time dependent tidal fields featuring a bar (top right panel), a bar with a density-wave spiral arm (bottom left panel) and a bar with transient spiral arms (bottom right panel). For comparison purposes, the $J_R$ and $L_Z$ range covered in the static potential case is shown as a box in the other panels and the actions of the Sun are marked in black.}
    \label{fig:amuse_actions}
\end{figure}

The actions of stars in the simulated clusters significantly help with our ability to identify solar siblings. The action distributions of the four simulations shown in Figure \ref{fig:amuse_positions} are displayed in Figure \ref{fig:amuse_actions}. As can be seen in the top-left panel of Figure \ref{fig:amuse_actions}, stars in the static potential remain centred around the current actions of the Sun ($J_{R,\odot}$,$L_{z,\odot}$,$J_{z,\odot}$). If the Sun's birth cluster was quite dense, then stars are even more tightly distributed around the Sun in action space. While it is not necessarily the case that the centre of mass of the Sun's birth cluster followed the Sun's exact orbit, Figure \ref{fig:amuse_actions} suggests that its initial orbital path could not have been far from the Sun's orbit in the static potential case as the action dispersion of escaped stars is narrow. More specifically, after combining stars in the low and high-density cluster simulations in a static tidal field, the dispersion in $J_R$ and $L_z$ are only $\sigma_{J_R,static}$ = 0.24 and $\sigma_{L_z,static}$ = 4.57 km $\rm s^{-1}$ kpc.

Similar to the $z$ distribution of solar siblings, the $J_z$ distribution of escaped stars in the static potential (not illustrated) is also very narrow. Each cluster's centre of mass maintains a constant $J_z$ for the entirety of the simulation, consistent with the solar value of $J_{z,\odot}$ = 0.37 km $\rm s^{-1}$ kpc. Hence escaped stars also have $J_z$ values near the solar value with a dispersion of $\sigma_{J_z,static}$ = 0.04 km $\rm s^{-1}$ kpc (ranging between 0.2 and 0.5 km $\rm s^{-1}$ kpc), due to star-star interactions. Depending on the initial conditions, stars can reach as low as 0.07 km $\rm s^{-1}$ kpc.

Unlike the spatial distribution of solar siblings, the bar has a clear effect on the action distribution of stars. More specifically the $J_R$ distribution significantly broadens and becomes bi-modal, with the standard deviation in $J_R$ increasing by a factor of 7.5 compared to the static case. Repeated interactions with the bar will perturb stars and alter the radial period of their orbits, however the distribution is still centred around the Sun's actions. The $L_z$ distribution of stars is only slightly broadened by the bar while the effect on $J_z$ is negligible.

Both the density wave spiral arm and transient spiral arms have dramatic effects on the distribution of actions of solar siblings. The $J_R$ distributions are widened, with most stars pushed to larger values of $J_R$. While the Sun is again near the centre of the distribution, the standard deviation in $J_R$ is over 10 and 100 times the static potential case for the density wave spiral arm and transient spiral arm models respectively. Similar effects are found for $L_z$ as well. Conversely, the $J_z$ distribution of stars again remains narrow (between 0.2 and 0.5 km $\rm s^{-1}$ kpc) for clusters orbiting in each galaxy model. The orbit of the model cluster's themselves undergo minor fluctuations in $J_z$, with a standard deviation about the solar value of only of 0.004 km $\rm s^{-1}$ kpc due to spiral arms. Only in a few extreme cases do interactions with the bar and spiral arms cause escaped stars to reach $J_z$ values up to 0.85 km $\rm s^{-1}$ kpc.

Given the results of Figures \ref{fig:amuse_positions} and \ref{fig:amuse_actions}, we can conclude that stars that may be far from the Sun spatially but have solar abundances and solar-like actions are potentially solar siblings. However it is still possible for stars with solar abundances that are far from the Sun, both spatially and in action-space, to be solar siblings due to interactions with the bar and spiral arms. Unfortunately, associating these stars with the Sun is difficult without exact knowledge of the properties and time evolution of the Milky Way and its various components.

\subsection{Solar Siblings in Analytic Potentials with Giant Molecular Clouds} \label{s_gmcs}

\begin{figure}
    \includegraphics[width=0.48\textwidth]{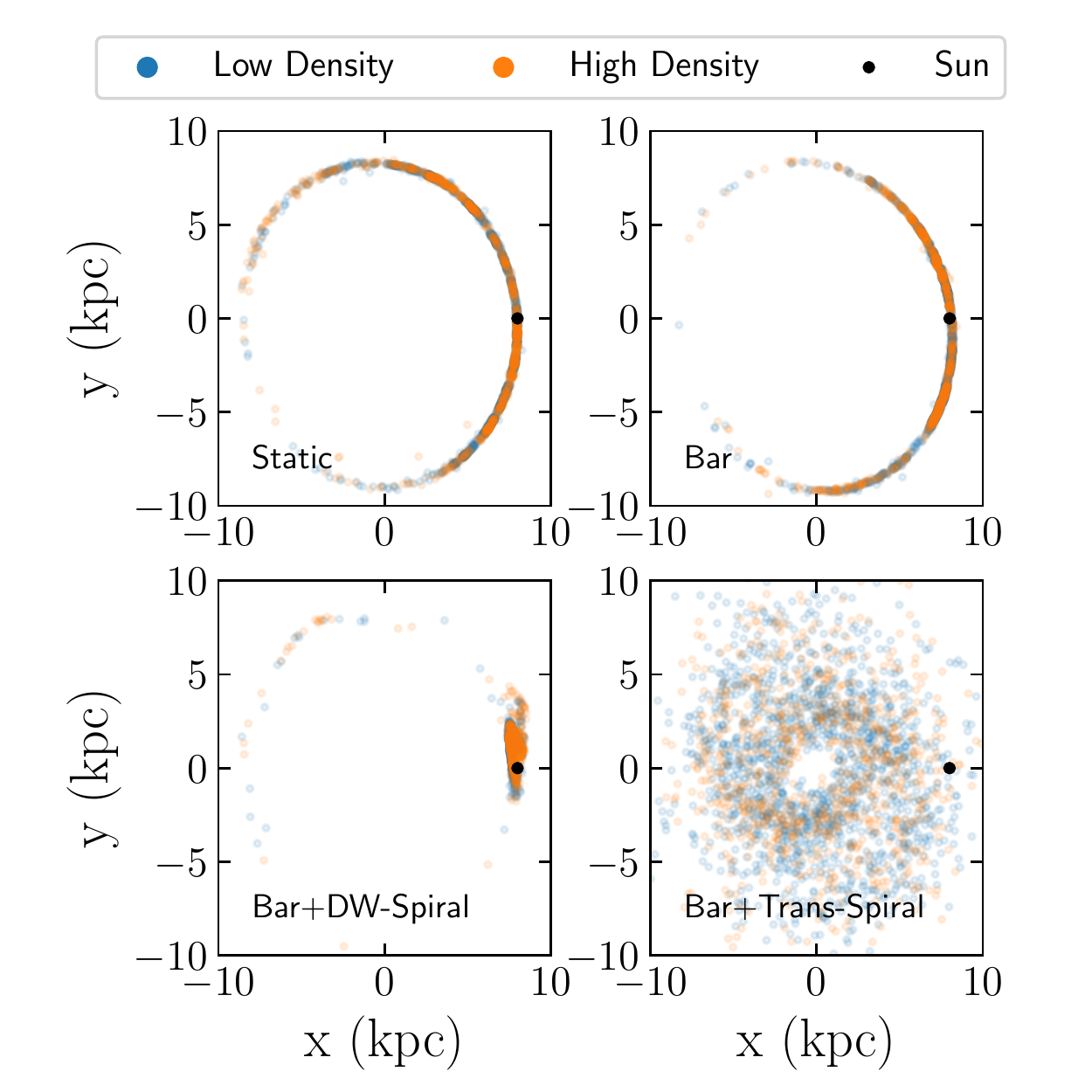}
    \caption{Same as Figure \ref{fig:amuse_positions} but for galaxy models containing GMCs.}
    \label{fig:amuse_positions_gmcs}
\end{figure}

\begin{figure}
    \includegraphics[width=0.48\textwidth]{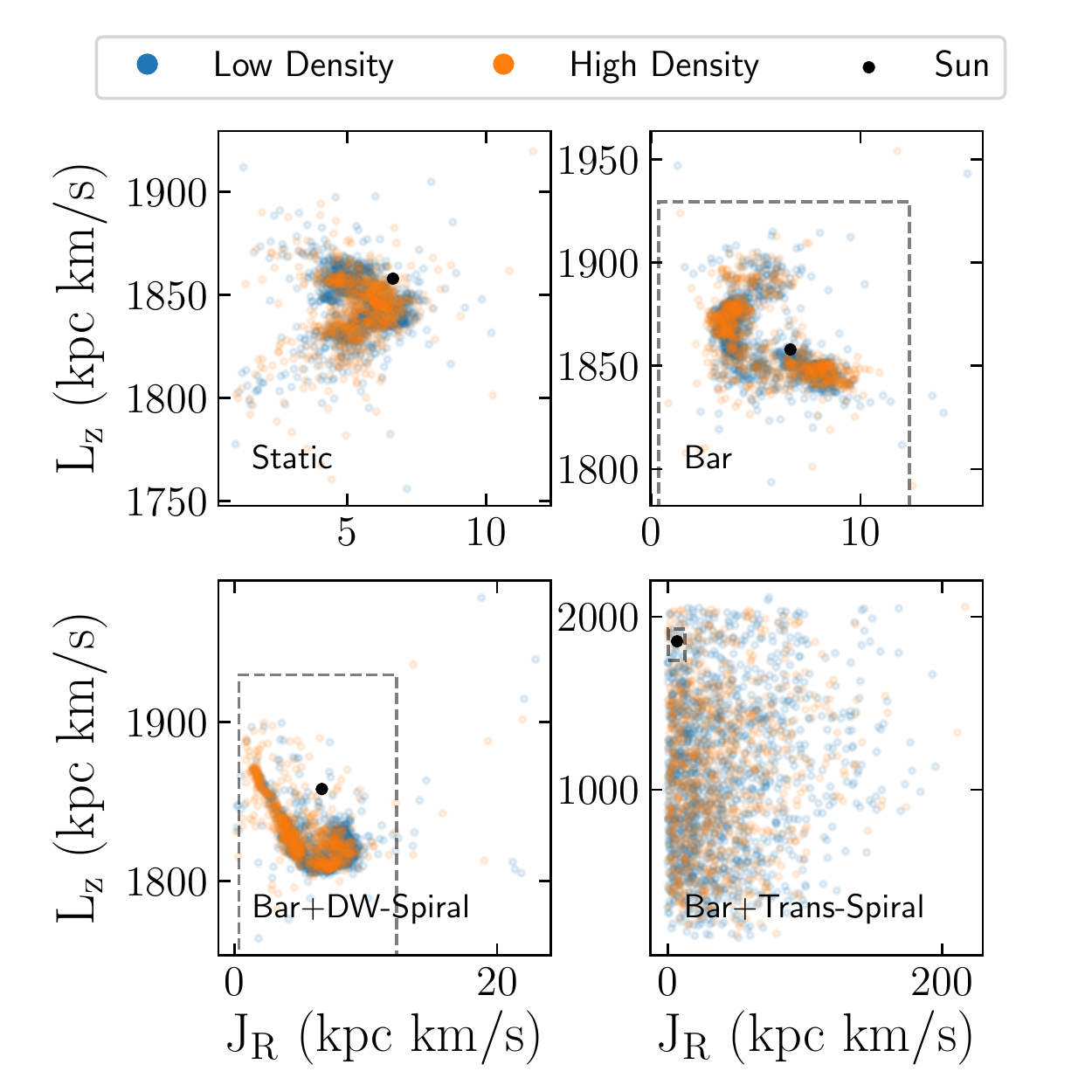}
    \caption{Same as Figure \ref{fig:amuse_actions} but for galaxy models containing GMCs.}
    \label{fig:amuse_actions_gmcs}
\end{figure}

Given how much we find time-dependent tidal fields can alter the spatial and action distributions of stars escaping a star cluster in Section \ref{s_pots}, it is worthwhile to consider another time-dependent element that is known to affect star cluster evolution. GMCs are known to be one of the primary sources of mass loss experienced by clusters \citep{Gieles2006, Kruijssen11} and have recently been found to even strongly influence the properties of tidal tails \citep{Banik19}. Therefore we explore the same four galaxy models discussed in Section \ref{s_pots}, but with the addition of a population of GMCs orbiting in the Galactic disc. The final positions of solar siblings are illustrated in Figure \ref{fig:amuse_positions_gmcs} and the actions of each star are shown in Figure \ref{fig:amuse_actions_gmcs}.

Figure \ref{fig:amuse_positions_gmcs} confirms the results of previous studies that GMCs accelerate cluster dissolution. In both of the top two panels it can be seen that the resulting stellar streams are both longer and thicker, signs of an accelerated mass loss rate. There also exists several gaps due to GMCs passing through or nearby part of the stream. While the transient spiral arm galaxy model with GMCs also follows this trend, the density spiral arm model surprisingly does not. 

The majority of stars in the bottom-left panel of \ref{fig:amuse_positions_gmcs} are still near the Sun's current position, despite a long low-density stream of stars that indicates GMCs have accelerated the cluster's dissolution rate. This apparent discrepancy can be explained by considering what orbit the star cluster must have to end up at the Sun's current position in the galaxy after 5 Gyr. In the galaxy model with a bar and density wave spiral arms, but no GMCs, the star cluster's orbit is initially quite eccentric with perigalactic and apogalactic distances of 7.5 kpc and 9.5 kpc. However with GMCs the cluster's orbit initially has a low eccentricity near 8.5 kpc, meaning it experiences a weaker tidal field and less tidal heating than the non-GMC model. Therefore it loses less mass due to tidal stripping. Conversely, adding GMCs to the transient spiral galaxy model means the progenitor cluster initially has a near-circular orbit at 4 kpc where many of the GMCs orbit before GMCs cause it to slowly migrate outward to the solar radius. Hence this cluster experiences a significant increase in mass loss due to both tidal stripping and GMC interactions. It is important to note that a different realization of the spiral arms and the GMC population may not produce such dramatically different orbital histories. 

Studying the distribution of solar sibling actions helps us quantify the effects of GMC interactions further. In the static, bar, and transient spiral arm potential models, interactions with GMCs slightly broaden the distributions of all three actions relative to the models that do not include GMCs. The effect is comparable to that adding just the bar to the static potential. Hence interactions with spiral arms are still the dominant mechanism behind altering the actions of stars that escape their host cluster. GMCs simply have the effect of slightly widening the range of actions that solar siblings could have. 

\begin{figure}
    \includegraphics[width=0.48\textwidth]{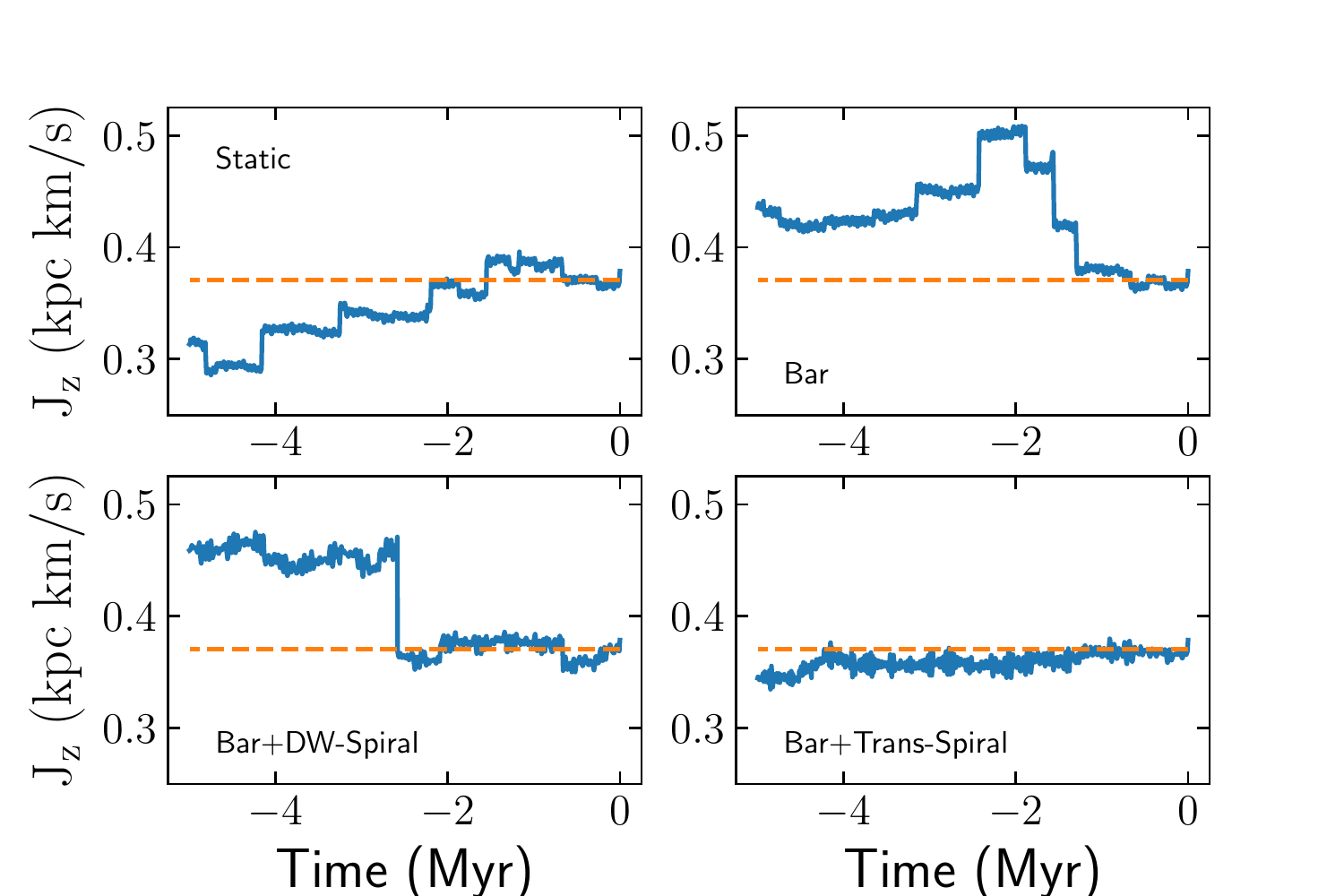}
    \caption{$J_z$ of the centre of mass of clusters in each galaxy model containing GMCs. The dashed horizontal line corresponds to the present day solar value of $J_z$.}
    \label{fig:amuse_jz_gmcs}
\end{figure}

Perhaps the most unique effect of GMCs, however, is that they increase the $J_z$ distribution of stars more than the bar or either spiral arm model. First, the orbital path of the model clusters themselves is unique due to interactions with GMCs. As illustrated in Figure \ref{fig:amuse_jz_gmcs}, $J_z$ of the cluster's centre of mass can vary between 0.3 and 0.5 km $\rm s^{-1}$ kpc before reaching the solar value at 5 Gyr. Hence stars that escape the cluster along the way will have a wider $J_z$ distribution compared to stars escaping a cluster with a fixed $J_z$. In all four cases, the dispersion in $J_z$ (again not illustrated) increases by roughly a factor 2 with stars having values between 0. and 2.2 km $\rm s^{-1}$ kpc. As we will show in the following section, expanding the search for solar siblings farther from the plane the disc increases the list of solar sibling candidates. However it should be noted that these results are sensitive to the initial setup of the GMC population, with different realizations and a prescription for GMC evolution potentially yielding different results if a cluster undergoes one or more close encounters.

\section{Matching Observed Actions to Simulations}\label{s_discussion}

Using accurately measured [Fe/H], [Mg/Fe], [Al/Fe], [Si/Fe], [K/Fe], [Ca/Fe] and [Ni/Fe] abundances, we have chemically tagged a list of giant stars in the APOGEE survey as solar sibling candidates. To narrow the list further, we make use of proper motions from \gaia to estimate the actions of each star. Actions are calculated assuming stars orbit in the \texttt{MWPotential2014} \citep{Bovy15} galaxy model. A large suite of simulations was then generated to help set the range of allowable actions that solar siblings can have. 


In order to determine which APOGEE stars with solar abundances have actions that fall within the same $J_R$ - $L_z$ and $J_z$ - $L_z$ parameter spaces as our simulations, we first combine all of the simulations into two datasets based on whether or not GMCs are included in the galaxy model or not. For each dataset, we define the allowable region of action space by generating a convex hull in both the $J_R$ - $L_z$ and $J_z$ - $L_z$ parameter spaces using the SciPy Python Pacakge \citep{SciPy}. In generating a convex hull, we identify a list of simulated stars that can be used to define the regions of $J_R$ - $L_z$ and $J_z$ - $L_z$ parameter space within which all simulated stars are found. Due to the sensitivity of convex hulls to outliers in the dataset, as the region must be expanded to incorporate all data points, it is necessary to exclude stars that are far removed from each action distribution due to a strong interaction with either a star in the cluster or a component of the Galaxy potential. Not doing so would result in the convex hull covering regions of action parameter space with low probabilities of actually having solar siblings. Therefore to ensure a conservative estimate of the allowed action space, we first calculate the difference between each star's actions and the mean actions of all stars in a given model. We then exclude the top $1\%$ of stars with the largest differences when generating the convex hulls. To incorporate the uncertainty in each action calculation due to the assumed Galactic potential, we also calculate actions assuming stars orbit in the Milky Way model proposed by \citet{McMillan17}. A star is considered a solar sibling candidate if its actions in either potential, corrected for how much $J_{R,\odot}$, $L_{z,\odot}$, and $J_{z,\odot}$ shift between the \texttt{MWPotential2014} and \citet{McMillan17} potentials, falls within each convex hull.

\begin{figure}
    \includegraphics[width=0.48\textwidth]{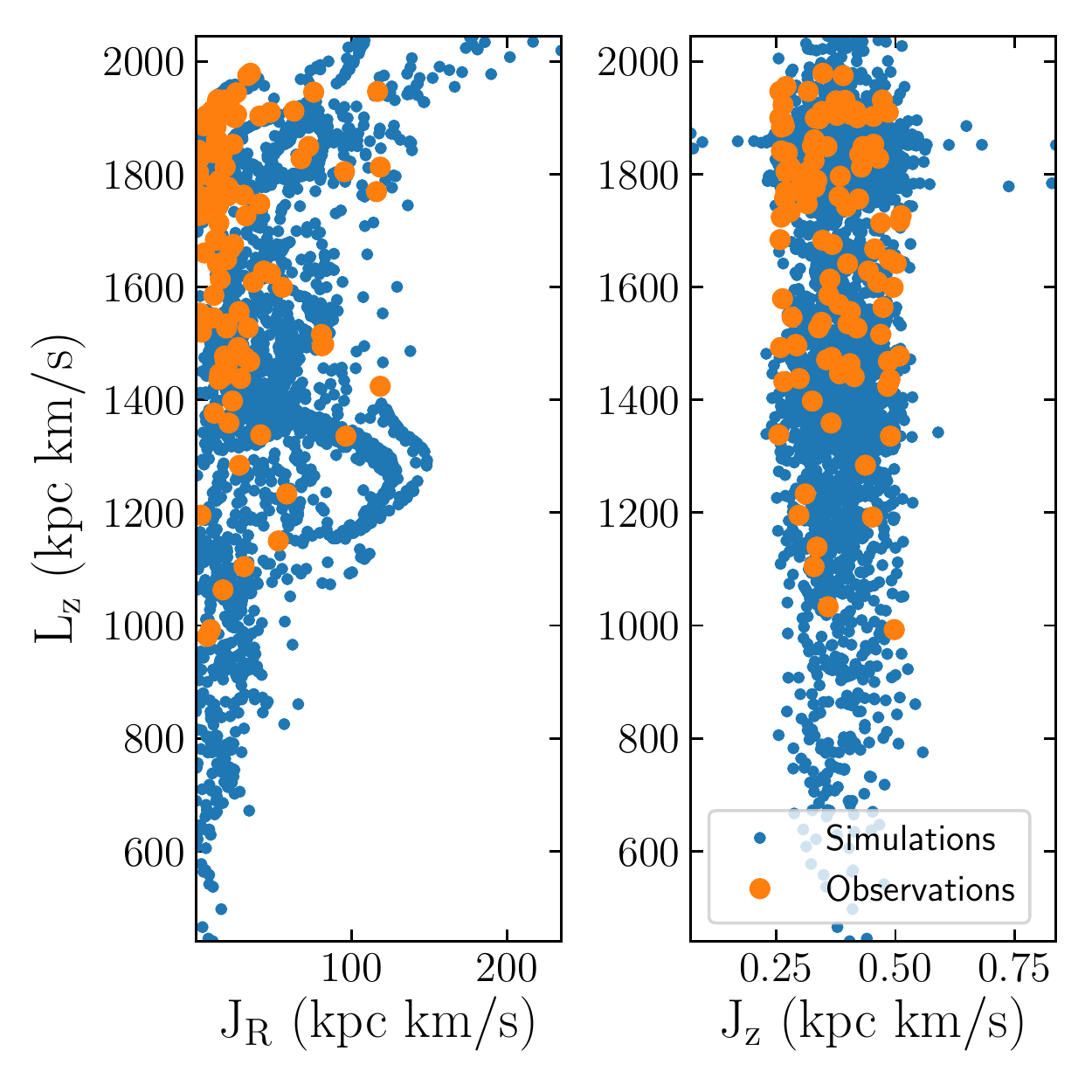}
    \caption{Distribution of actions $J_R$, $L_z$, and $J_z$ for stars in the low and high density model clusters orbiting in all four of the model tidal fields without GMCs (blue). 104 APOGEE stars with actions within the convex hull of the model clusters and abundances consistent with solar values for all considered elements (orange).}
    \label{fig:amuse_full_overlap}
\end{figure}

Ignoring, for the moment, simulations that include GMCs, we find a total of 104 APOGEE stars that have solar abundances and actions that fall within the range set by stars in the time-dependent galaxy models. These stars are considered to be primary candidates as specific interactions with GMCs are not required to explain their connection to the Sun. The action distributions of these stars are compared to the simulations in Figure \ref{fig:amuse_full_overlap}.

The range of actions spanned by the 104 primary candidates illustrated in Figure \ref{fig:amuse_full_overlap} are $J_R < 122$ km $\rm s^{-1}$ kpc, $990 < L_z < 1986$ km $\rm s^{-1}$ kpc, and $0.15 < J_z < 0.58$ km $\rm s^{-1}$ kpc. In order to rank our list of candidates we determine each stars distance in $J_R$, $L_z$, and $J_z$ from the solar values, where individual distances are normalized by $\sigma_{J_R,static}$, $\sigma_{L_z,static}$, and $\sigma_{J_z,static}$ respectively. Hence the parameter $\Delta \sigma_J$ is calculated via:

\begin{equation}\label{eqn:dsigma}
    \begin{split}
        \Delta \sigma_J^2 &= (J_R-J_{R,\odot})/\sigma_{J_R,static}^2+(L_z-L_{z,\odot})/\sigma_{L_z,static}^2+\\
    &(J_z-J_{z,\odot})/\sigma_{J_z,static}^2
    \end{split}
\end{equation}

\noindent Since $\Delta \sigma_J^2$ is normalized by $\sigma_{J_R,static}$, $\sigma_{L_z,static}$, and $\sigma_{J_z,static}$, it provides an indication of how close an observed star's actions are to the action distributions of stars simulated in the static galaxy model. Hence stars with low values of $\Delta \sigma_J^2$ depend less on the non-axisymmetric components of the Milky Way to be considered solar siblings. However it is important to note that $\Delta \sigma_J$ alone cannot be used to constrain solar siblings, as it does not take into consideration the three dimensional shape of the allowable $J_R$, $L_z$, and $J_z$ distribution. The parameter is only used as a measure of how close a candidate star's actions are to the Sun after it has been found to have solar abundances and overlaps with the simulated solar sibling dataset.

The candidate star with the lowest $\Delta \sigma_J$, which we will refer to as Solar Sibling 1 (SS1), has $\Delta \sigma_J$ = 5. For reference purposes, it is worth noting that escaped stars from simulated clusters in static galaxy models can reach up to $\Delta \sigma_J$ = 7.75. SS1 does not, however, directly overlap with simulated stars in the static galaxy models as its $L_z$ is slightly too low. The properties of SS1 are listed in Table \ref{table:matches}. The top five solar sibling candidates (SS1-SS5) have $\Delta \sigma_J$ < 14 or roughly twice the range covered by simulated clusters in static galaxy models.

\begin{table}
\centering
\begin{tabular}{lccc}
\hline
Variable & Solar Sibling 1 & Sun \\
\hline
APOGEE ID & 2M19354742+4803549 & {} \\
\hline
RA & {293.95$^\circ$} & {-} \\
Dec & {48.06$^\circ$} & {-}\\
Distance & {0.36 $\pm$ 0.08 kpc} & {-}\\
$\mu_{RA}$ & {2.85 $\pm$ 0.04 mas yr$^{-1}$} & {-} \\
$\mu_{Dec}$ & {2.12 $\pm$ 0.04 mas yr$^{-1}$} & {-} \\
$v_{los}$ & {-11.19 $\pm$ 0.02 km s$^{-s}$} & {-} \\
$J_R$ & {7.37 km $\rm s^{-1}$ kpc} & {6.65 km $\rm s^{-1}$ kpc} \\
$L_z$ & {1865.78 km $\rm s^{-1}$ kpc} & {1857.92 km $\rm s^{-1}$ kpc} \\
$J_z$ & {0.41 km $\rm s^{-1}$ kpc} & {0.37 km $\rm s^{-1}$ kpc} \\
\hline
{[Fe/H]} & {0.02 $\pm$ 0.04} & 0.066 $\pm$ 0.009 \\
{[Mg/Fe]} & {-0.02 $\pm$ 0.06}  & 0.010 $\pm$ 0.002 \\
{[Al/Fe]} & {-0.1 $\pm$ 0.1} & -0.012 $\pm$ 0.007 \\
{[Si/Fe]} & {-0.07 $\pm$ 0.07} & -0.011 $\pm$ 0.004 \\
{[K/Fe]} & {-0.01 $\pm$ 0.06} & -0.013 $\pm$ 0.004 \\
{[Ca/Fe]} & {0.02 $\pm$ 0.05}  & -0.008 $\pm$ 0.003 \\
{[Ni/Fe]} & {-0.03 $\pm$ 0.06} & 0.015 $\pm$ 0.002 \\
\hline 

\end{tabular}
\caption{\label{table:matches} Orbital (top half) and chemical (lower half) properties of Solar Sibling 1 and the Sun.}
\end{table}

\begin{figure}
    \includegraphics[width=0.48\textwidth]{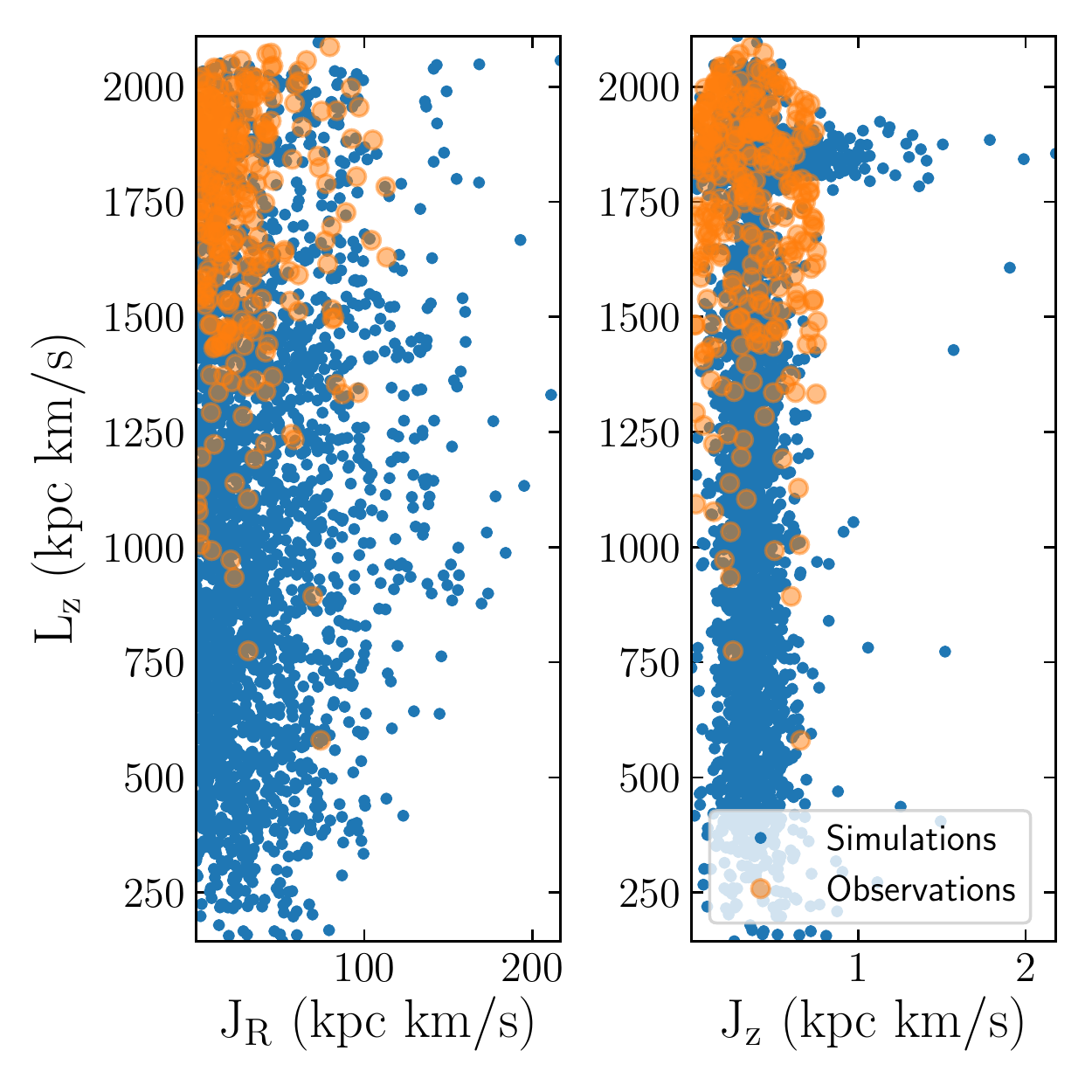}
    \caption{Same as Figure \ref{fig:amuse_full_overlap}, but for model tidal fields that include GMCs. 296 stars in APOGEE have solar abundances and actions within the range of simulations including GMCs.}
    \label{fig:amuse_full_overlap_gmcs}
\end{figure}

As discussed in Section \ref{s_gmcs}, including GMCs in our simulations broadens the  $J_R$, $L_z$, and $J_z$ distributions of escaping stars. Furthermore, GMC interactions are the primary mechanism for increasing the $J_z$ distribution of stars. Hence we expect the list of potential candidates to expand when incorporating the effects of GMCs, which is exactly what is observed in Figure \ref{fig:amuse_full_overlap_gmcs}.

An additional 192 stars (SS105-SS296) overlap in action-space with clusters evolved in tidal fields that include GMCs. The largest $\Delta \sigma_J$ reached by solar sibling candidates that overlap with the simulated clusters in galaxy models that include GMCs is 296. Combining the entire suite of simulations allows for the total allowable region of action space to be constrained to $J_R < 122$ km $\rm s^{-1}$ kpc, $353 < L_z < 2110$ km $\rm s^{-1}$ kpc, and $J_z < 0.8$ km $\rm s^{-1}$ kpc. While the $J_z$ distribution of secondary candidates still primarily lies near the solar value of $\sim$ 0.37 km $\rm s^{-1}$ kpc when including GMCs, GMC interactions lead to stars being scattered to higher and lower values of $J_z$. 

It is still difficult to rule out APOGEE stars with solar abundances that we do not consider candidates, as the exact details of the Milky Way's tidal field and its history (combined with GMC interactions) could further widen the allowed action distribution of solar siblings. It is for this reason that we refer to stars that overlap in action space with simulations in time-dependent tidal fields as primary candidates when GMCs are included and secondary candidates when GMCs are not included. A table listing the properties of all primary and secondary candidates can also be found in the on-line catalogue, in order of increasing $\Delta \sigma_J$.

Primary candidates with low values of $\Delta \sigma_J$ are less dependent on the exact details of the Milky Way's gravitational field, as strong interactions with the Galatic bar or spiral arms are not necessary to explain the difference between their actions and the solar values. Primary candidates with larger values of $\Delta \sigma_J$, however, depend on our assumed properties of the Galactic bar and spiral arms. A bar that has a resonance, for example, with the Sun's birth cluster may lead to a wider distribution of solar sibling actions than the bar model considered here. Secondary candidates further depend on the exact cluster-GMC interactions that occur within our simulations and our simplified treatment of GMC evolution (circular orbits, fixed masses, and fixed sizes), both of which could vary from one realization to the next. Both candidate lists are limited by the fact that we only consider a single cluster orbit, manufactured so that its centre of mass has a solar orbit at the end of the simulation. Simulating several realizations of clusters with a range of different orbits, while also incorporating a prescription for GMC evolution, will likely increase the list of candidates as clusters will be subjected to a wider range of bar, arm, and GMC interactions. 

\subsection{Comparison to Previous Solar Sibling Candidates}

The volume of action space that constrains stars in the entire suite of simulations can be applied to previously known solar sibling candidates to strengthen or rule out their candidacy.  As previously discussed, \citet{Martinez16} compared the spatial and kinematic properties of 31 known solar sibling candidate to their own suite of simulations and found none of the candidates had a high probability of being born in the same cluster as the Sun. HD 147443 and HD 196676 were found to be spatially and kinematically consistent with being solar siblings, but neither star has solar chemical compositions despite being comparable in age and metallicity to the Sun. Applying our action space criteria to the remaining stars in the \citet{Martinez16}  dataset, we find 4 stars (HD175740, HD52242, HD83423, HD168442) that would be considered primary candidates and a further 13 (HD192324 'HD46301, HD26690, HD207164, HD105678, HD28676, HD95915, HD105000, HD44821, HD199951, HD168769, HD46100, HD154747), that would be considered to be secondary candidates if they also have solar abundances. The majority of the 12 stars that we rule out have $J_z$ values that are too high when compared to our simulations, while the other 3 have $L_z$ values that are too high.


The actions of the recently discovered solar twin HD186302 \citep{Adibekyan18} also do not meet the action phase space criteria of being a solar sibling. Specifically its $J_z$ is $\sim$ 1.3 km $\rm s^{-1}$ kpc, which is outside the $J_z$ range reached by $99\%$ of the simulated stars. It is worth noting, however, that in the galaxy models with GMCs a small number of stars are scattered to values of $J_z$ of $\sim$ 2 km $\rm s^{-1}$ kpc. Hence, while we cannot rule out HD186302 as a solar sibling, it less likely to be a solar sibling than any of the primary or secondary candidates presented here. A close GMC interaction is required to push solar siblings to such high $J_z$ values.

\subsection{Comparison to M67}

Since the suite of simulations presented above are focused on progenitor clusters that end up at the Sun's position after 5 Gyr, they by design do not produce any models where the Sun's birth cluster, or its remnant, ends up near the current location of M67. To address the issue of whether or not M67 could be the Sun's birth cluster, we re-simulate the low and high-density model clusters in the galaxy models with GMCs on orbits that put them at the present day position of M67 after 5 Gyr \citep{Loktin03, Xin05, Conrad17}. Only the galaxy models with GMCs were considered as \citet{Gustafsson16} finds that GMC interactions are required for a star to escape M67 on a solar orbit and then have the cluster migrate to its current position.

We find that only the $J_R$ values of stars that escape our model M67 are consistent with being solar, while some stars in the galaxy models with spiral arms also have $L_z$ values that are near solar. However the $J_z$ values of both M67 and stars that have escaped the cluster over the past 5 Gyr are approximately ten times larger than that of the Sun. Hence our simulations indicate that it is unlikely the Sun was born in M67, in agreement with \citep{Pichardo12}. The only way that the Sun could have been formed in M67 would be if M67 has undergone one or more very energetic GMC encounters to push the cluster to its current $J_z$. \cite{Jorgensen19} found that the probability of such an event occurring \textit{and} not destroying M67 completely is low.

\section{Conclusion}\label{s_conclusion}

We construct a list of stars in APOGEE/\gaia\ whose abundances and kinematics are consistent with them once being members of the same host cluster as the Sun. Stars were first selected as potential solar sibling candidates based on their chemical abundances. More specifically, any star with values of [Fe/H[, [Mg/Fe], [Al/Fe], [Si/Fe], [K/Fe], [Ca/Fe], and [Ni/Fe] that were consistent with being solar within their measurement uncertainty were chosen to be potential candidates. To constrain the list of solar sibling candidates further, we compared their actions to a suite of star clusters simulations.

Star clusters were evolved in a range of static and time-dependent fields such that they have dissolved and the progenitor remnant has reached the Sun's current position in the Milky Way. The actions of each star were then solved assuming a static tidal field for the Galaxy model. The static potential consists of a bulge, disk and halo while time-dependent features like a bar, spiral arms, and GMC encounters were added separately and together. 

The simulations indicate that in a static tidal field, the range of actions that stars which escape their host cluster can have is rather narrow ($5.8 < J_R < 7.4$ km $\rm s^{-1}$ kpc, $1848. < L_z < 1868$ km $\rm s^{-1}$ kpc, and $0.27 < J_z < 0.49$ km $\rm s^{-1}$ kpc.). Including a bar and spiral arms can broaden the $J_R$ and $L_z$ distribution of escaped stars, but leaves the $J_z$ distribution untouched. Interactions with spiral arms, specifically transient spiral arms, are particularly strong and lead to the widest distribution of actions that stars from the same birth cluster can have. The wide spatial and kinematic distributions of model stars due to interactions spiral arms highlights the importance of chemical tagging when searching for stars that formed in the same cluster. Interactions with GMCs can slightly broaden the $J_R$ and $L_z$ distributions further, while also having the unique effect of widening the $J_z$ distribution of escapes stars as well. Hence it is possible for escaped stars to exist beyond the plane of the disk. Taking into consideration the entire suite of simulations, we find that solar siblings are most likely to have $J_R < 122$ km $\rm s^{-1}$ kpc, $353 < L_z < 2110$ km $\rm s^{-1}$ kpc, and $J_z < 0.8$ km $\rm s^{-1}$ kpc.

In total we find 296 stars in APOGEE with solar-like abundances that also have actions that overlap with our suite of simulations. The list of candidates can be broken up into primary and secondary lists based on whether or not the model stars that the candidates have similar actions to were evolved in a time-dependent field with no GMCs or a time-dependent field with GMCs. Based on these criteria, we find our list of solar sibling candidates consists of 104 primary candidates and 192 secondary candidates. 


The top primary candidate, SS1, is the most promising solar sibling candidate as its actions are quite close to the solar values. In fact its actions nearly overlap with simulated stars in the static galaxy model, indicating that it does not specifically require strong interactions with the bar, spiral arms, or GMCs in order to be associated with the Sun's birth cluster. Candidates SS2-SS5 are also similarly close to the distribution of escapers in the static galaxy model. The inclusion or exclusion of stars as primary or secondary candidates is sensitive to the assumed details of the Milky Way's bar, spiral arms, and GMC population. Hence we cannot conclude that the primary or secondary candidates are solar siblings or that the excluded stars with solar-like abundances are not. It is further likely that the candidate list can be expanded further by exploring how different initial cluster orbits as well as different bar, spiral arm, and GMC models affect the distribution of simulated solar siblings. The list of candidates presented here should, therefore, be considered conservative. It is worth noting, however, that the farther the actions of a given star are from the distributions of our simulated clusters, the less likely it is of being a solar sibling.

We also re-examine several solar sibling candidates from past studies \citep{Martinez16,Abolfathi2018} and find 17/30 have actions that meet our criteria as either primary or secondary candidates. $J_z$ is typically the action that is responsible for these stars not meeting our criteria, which indicates they could only be a solar sibling if the Sun's birth cluster underwent a strong interaction with a GMC.

Finally, we also consider the possibility of M67 being the Sun's birth cluster by re-simulating model clusters on orbits that have them reach the present day position of M67 after 5 Gyr. In no scenario does an escaped star have comparable actions to that of the Sun, primarily due to the very large $J_z$ value associated with M67's orbit. Even encounters with GMCs are not enough to allow a star to escape M67 on a solar orbit and then push the birth cluster to such a large $J_z$. Since we only consider one realization of a GMC population here, we can only conclude that the probability of M67 being the Sun's birth cluster is low and that a very strong GMC encounter would be required to account for the large $J_z$ difference. Complicating the scenario further is the fact that the strong encounter must not cause the cluster to completely disrupt.

The combined approach of chemical tagging and action analysis offers a strong method for constraining not only solar siblings, but any group of stars that formed from the same GMC. Applying the above method to large datasets will allow for constraints to be placed on star and star cluster formation, the assembly history of the Milky Way, and the origin of the Sun. A complete method will have to account of variations in the properties of the assumed Galactic potential and GMC population, which will likely broaden the allowed volume of action space that stellar siblings can populate. A large sampling of initial cluster properties and orbits will also help apply probabilities to sibling candidates and provide an indication of the frequency of false positives. Confirmation of candidates as solar siblings would require follow-up observations in the form of high-resolution optical spectroscopy to constrain more abundances, in particular different r- and s-process elements, and asteroseismology to accurately measure ages.

\section*{Acknowledgements}

JW acknowledges financial support through a Natural Sciences and Engineering Research Council of Canada (NSERC) Postdoctoral Fellowship. NPJ is supported by an NSERC Alexander Graham Bell Canada Graduate Scholarship-Doctoral. JW, NPJ, and JB also acknowledge additional financial support from NSERC (funding reference number RGPIN-2015-05235) and an Ontario Early Researcher Award (ER16-12-061). SPZ thanks Norm Murray and the Canadian Institute for Theoretical Astrophysics (CITA) for the hospitality during his long-term visit. JH is supported by a Dunlap Fellowship at the Dunlap Institute for Astronomy \& Astrophysics, funded through an endowment established by the Dunlap family and the University of Toronto and the Flatiron Institute, which is supported by the Simons Foundation.

Funding for the Sloan Digital Sky Survey IV has been provided by the Alfred P. Sloan Foundation, the U.S. Department of Energy Office of Science, and the Participating Institutions. SDSS-IV acknowledges
support and resources from the Center for High-Performance Computing at
the University of Utah. The SDSS web site is \url{www.sdss.org}.





\bibliographystyle{mnras}
\bibliography{main} 


\bsp	
\label{lastpage}
\end{document}